\documentclass[aps, prl, reprint, superscriptaddress, floatfix]{revtex4-2}
\usepackage[english]{babel}
\usepackage[utf8]{inputenc}
\usepackage[T1]{fontenc}
\usepackage{hyperref, xcolor, graphicx}
\usepackage{amsfonts, amssymb, amsmath, mathtools}

\newcommand\beq{\begin{equation}}
\newcommand\eeq{\end{equation}}
\newcommand{\pd}{\partial}

\hypersetup{
    colorlinks=true,
    linkcolor=black,
    citecolor=blue,   
    urlcolor=blue,
}

\begin{document}

\title{Quantum chaos without false positives}

\author{Dmitrii~A.~Trunin}
    \email{dmitriy.trunin@phystech.edu}
    \affiliation{Moscow Institute of Physics and Technology, 141701, Institutskiy pereulok 9, Dolgoprudny, Russia}
    \affiliation{Lebedev Physical Institute, 119991, Leninskiy prospect 53, Moscow, Russia}

\date{\today}

\begin{abstract}
Out-of-time-order correlators are widely used as an indicator of quantum chaos, but give false-positive quantum Lyapunov exponents for integrable systems with isolated saddle points. We propose an alternative indicator that fixes this drawback and retains all advantages of out-of-time-order correlators. In particular, the new indicator correctly predicts the average Lyapunov exponent and the Ehrenfest time in the semiclassical limit, can be calculated analytically using the replica trick, and satisfies the bound on chaos.
\end{abstract}

\maketitle

\textit{Introduction}.-- Out-of-time-order correlators (OTOCs) enjoy an exceptionally wide range of applications from condensed matter physics to quantum
gravity. The most important application of OTOCs is the identification of quantum chaos. This application stems from a generalization of the Lyapunov exponent (LE), which measures the exponential sensitivity to initial conditions in classical chaotic systems~\cite{Larkin, MSS, Kitaev-talks, Swingle-popular, Garcia-Mata:2022}. Namely, to extend this property to quantum systems, we rewrite the sensitivity using a Poisson bracket~\footnote{We consider a Hamiltonian system and introduce the canonical coordinates on the phase space $\mathbf{z} = (\mathbf{q}, \mathbf{p})$.}, $\pd z_i(t) / \pd z_j(0) = \left\{ z_i(t), z_j(0) \right\}$, quantize it, average over a thermal ensemble, and define the OTOC $C(t)$ and the quantum LE $\kappa_q$:
\beq \label{eq:OTOC-def}
\begin{aligned}
C(t) &= \sum_{i,j} \left\langle \left[ \hat{z}_i(t), \hat{z}_j(0) \right]^\dag \left[ \hat{z}_i(t), \hat{z}_j(0) \right] \right\rangle \sim e^{2 \kappa_q t}.
\end{aligned} \eeq
This definition implies that classical chaotic systems acquire a positive quantum LE upon quantization, so quantum chaos is naturally associated with $\kappa_q > 0$.

Furthermore, this definition of quantum chaos is straightforwardly extended to quantum many-body systems and proves to be related to thermalization and information scrambling~\cite{Shenker-1, Shenker-2, Roberts-1, Shenker-3, Swingle:2016, Roberts:2016, Nahum, Mi, Xu:2018, Xu-tutorial}. The latter property is especially important for large-$N$ quantum systems holographically dual to black holes. Indeed, black holes are the fastest scramblers in nature~\cite{Hayden, Sekino, Lashkari} and impose a bound on the quantum LE~\cite{MSS}. Hence, if a quantum system saturates the bound, it is likely dual to a black hole and provides a qualitative model of its microstates, which opens a way for experimental simulation of black holes~\cite{Garcia-Alvarez, Jafferis:Nature}. Besides, OTOCs are relatively easy to calculate analytically and measure experimentally~\cite{Garttner, Li, Green}. This makes the OTOCs an indispensable tool and explains the ever-growing interest in them in both the condensed matter and high-energy physics communities~\cite{Polchinski, Maldacena-SYK, Kitaev, Sarosi, Rosenhaus, Trunin-SYK, Maldacena-JT, Jensen, Engelsoy, Gross-1, Gross-2, Roberts-2, Fitzpatrick, Turiaci, Stanford:phi-4, Grozdanov, Hashimoto:2017, Akutagawa, Kolganov, Buividovich:2018, Buividovich:2022, Bohrdt, Klug, Swingle:2017, Patel, Tikhanovskaya, Kent:2023, Rozenbaum-1, Rozenbaum-2, Xu, Pilatowsky-Cameo:2019, Wang:2018, Pappalardi:2018, Hummel:2018, Hashimoto:2020, Steinhuber:2023, Dowling:2023}.

Nevertheless, OTOCs have a serious drawback: they grow exponentially in classically integrable systems with isolated saddle points~\cite{Rozenbaum-1, Rozenbaum-2, Xu, Pilatowsky-Cameo:2019, Wang:2018, Pappalardi:2018, Hummel:2018, Hashimoto:2020, Steinhuber:2023, Dowling:2023}. The primary source of such a false growth is the incorrect order of averaging over the phase space and taking logarithm in the definition of the quantum LE, which magnifies the contribution of nontypical exponentially diverging trajectories in a small vicinity of a saddle point. So, such false positives call into question the use of OTOCs as an indicator of chaos in quantum systems with a well-defined classical limit.

To close this loophole, we suggest an alternative indicator of quantum chaos, which we refer to as the logarithmic OTOC (LOTOC)~\footnote{Note that the operator under logarithm is Hermitian and positive definite, so the logarithm is well defined.}:
\beq \label{eq:LOTOC-def}
L(t) = \left\langle \log\!\bigg( \sum_{i,j} \left[ \hat{z}_i(t), \hat{z}_j(0) \right]^\dag \left[ \hat{z}_i(t), \hat{z}_j(0) \right] \bigg) \right\rangle. \eeq
The refined quantum LE $\bar \kappa_q$ is extracted from the linear growth of LOTOC up to the Ehrenfest time~\cite{Shepelyansky:Scholarpedia, Zaslavsky, Chirikov-88, Aleiner-96}, where semiclassical description fails and $L(t)$ saturates:
\beq \label{eq:LLE-def}
L(t) \approx 2 \bar \kappa_q t + o(t), \quad 1 \ll t \ll t_E. \eeq
Here, $o(t)$ grows slower than linearly (e.g., $o(t) \sim \log t$).

In this Letter, we argue that the refined quantum LE coincides with the phase-space average of the classical LE in the semiclassical limit. This allows us to reliably distinguish between chaotic ($\bar \kappa_q > 0$) and integrable ($\bar \kappa_q = 0$) quantum systems, including systems with isolated saddle points. In a sense, our definition of quantum chaos generalizes the definition of a Kolmogorov system with positive Kolmogorov-Sinai entropy~\cite{Tabor, Pesin}.

Moreover, we show that the LOTOC retains the most important advantages of the conventional OTOC: it can be calculated analytically in the large-$N$ systems and satisfies the bound on chaos~\cite{MSS}. To this end, we rewrite the logarithm in definition~\eqref{eq:LOTOC-def} using the replica trick:
\beq \label{eq:replica-trick}
L(t) = \lim_{n \to 0} \frac{\pd C_n(t)}{\pd n} \qquad \text{and} \qquad \bar{\kappa}_q = \lim_{n \to 0} \frac{\pd \kappa_n}{\pd n}, \eeq
where we introduce the replica OTOC:
\beq \label{eq:ROTOC}
C_n(t) = \left\langle \bigg( \sum_{i,j} \left[ \hat{z}_i(t), \hat{z}_j(0) \right]^\dag \left[ \hat{z}_i(t), \hat{z}_j(0) \right] \bigg)^n \right\rangle,
\eeq
and the replica LE $\kappa_n$:
\beq C_n(t) = 2 \kappa_n t + o(t), \quad 1 \ll t \ll t_E. \eeq
Similarly to OTOCs, which are naturally defined using the two-fold Keldysh contour~\cite{Aleiner, Haehl, Romero-Bermudez}, replica OTOCs are conveniently calculated using the Schwinger-Keldysh diagram technique on a $2n$-fold contour. Furthermore, in the \mbox{large-$N$} limit, which is most interesting in the context of holography, the behavior of correlators on the $2n$-fold contour becomes rather distinguished, so the replica OTOCs can be estimated analytically. In the following, we present several examples of such a calculation. For more details on the extended Schwinger-Keldysh technique and calculation of replica OTOCs, see~\cite{companion}.

\textit{False chaos}.-- As an illustrative example of an integrable system with an isolated saddle point, we consider the Lipkin-Meshkov-Glick (LMG) model~\cite{Xu, Pilatowsky-Cameo:2019, Wang:2018, Pappalardi:2018, LMG-1, *LMG-2, *LMG-3}:
\beq \label{eq:LMG}
\hat H_\mathrm{LMG} = \hat{x} + 2 \hat{z}^2, \eeq
where $\hat{x}, \hat{y}, \hat{z} = \hat{S}_x/S, \hat{S}_y/S, \hat{S}_z/S$ are rescaled $SU(2)$ spin operators with total spin $S$. In the classical limit $S \to \infty$, these operators form a classical $SU(2)$ spin that lives on a unit sphere $x^2 + y^2 + z^2 = 1$, and the commutation relation $[ \hat{x}_m, \hat{x}_n] = i \hbar \epsilon_{mnk} \hat{x}_k$ with the effective Planck constant $\hbar = 1/S$ is replaced by the corresponding Poisson bracket, $\{ x_m, x_n \} =  \epsilon_{mnk} x_k$. The phase space of the classical LMG model has dimension two, so it is automatically integrable. At the same time, this model has an isolated saddle point $x = 1$, where $\pd z_i(t) / \pd z_j(0) \sim e^{\kappa_s t}$ with $\kappa_s = \sqrt{3}$.

\begin{figure}
    \centering
    \includegraphics[width=\linewidth]{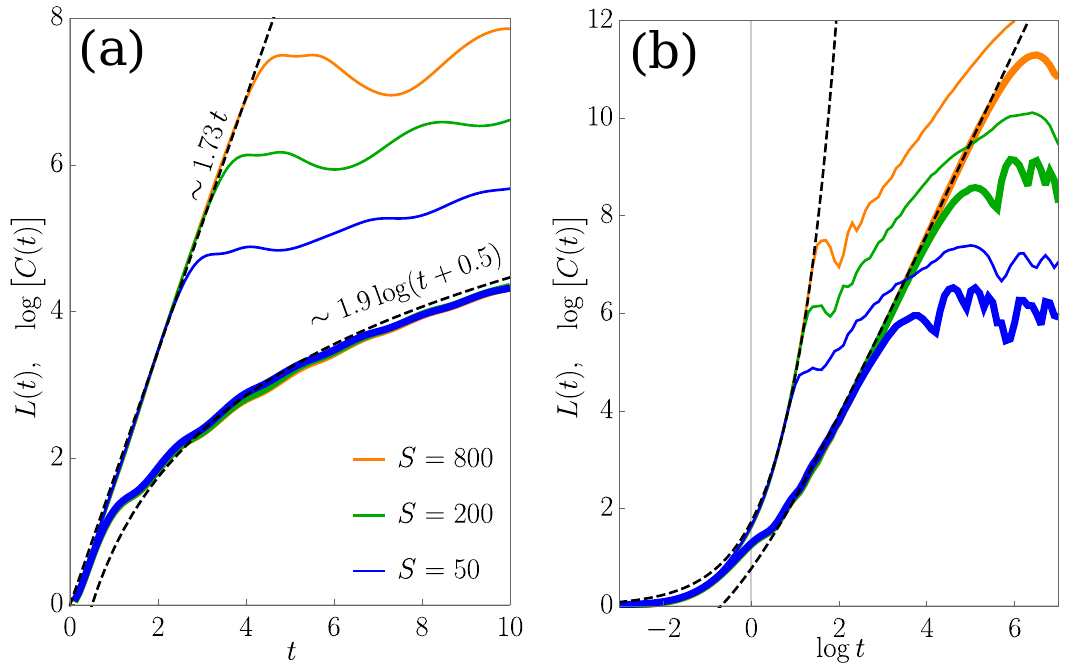}
    \caption{(a) Infinite-temperature OTOCs (thin lines) and LOTOCs (thick lines) in the integrable LMG model. (b) The same plot in the logarithmic timescale.}
    \label{fig:LMG}
\end{figure}

Let us calculate the OTOC and the LOTOC in the model~\eqref{eq:LMG}. For simplicity, we parametrize the phase space using $(x,y,z)$ coordinates~\footnote{On a K{\"a}hler manifold, classical LEs do not depend on the space parametrization, and we believe this property to be preserved after the semiclassical quantization.} and consider the infinite-temperature limit, where the behavior of correlation functions~\eqref{eq:OTOC-def} and~\eqref{eq:LOTOC-def} is most pronounced. The numerical result, Fig.~\ref{fig:LMG}, shows that the OTOC grows exponentially up to the ``chaotic'' Ehrenfest time, $C(t) \sim e^{2 \kappa_q t}$ for $1 \lesssim t \lesssim \log(1/\hbar)$; furthermore, $\kappa_q \approx \kappa_s / 2$. On the contrary, the LOTOC grows logarithmically until it saturates at much larger ``integrable'' Ehrenfest time, $L(t) \sim \log t$ for $1 \lesssim t \lesssim 1/\hbar$. Hence, the definition~\eqref{eq:LLE-def} implies that the refined quantum LE is zero, as it should be in an integrable system. Moreover, this behavior indicates that the semiclassical dynamics of a quantized integrable system is correcly captured by the LOTOC rather than the OTOC (also compare with~\cite{Pappalardi:2018}).

\begin{figure}
    \centering
    \includegraphics[width=\linewidth]{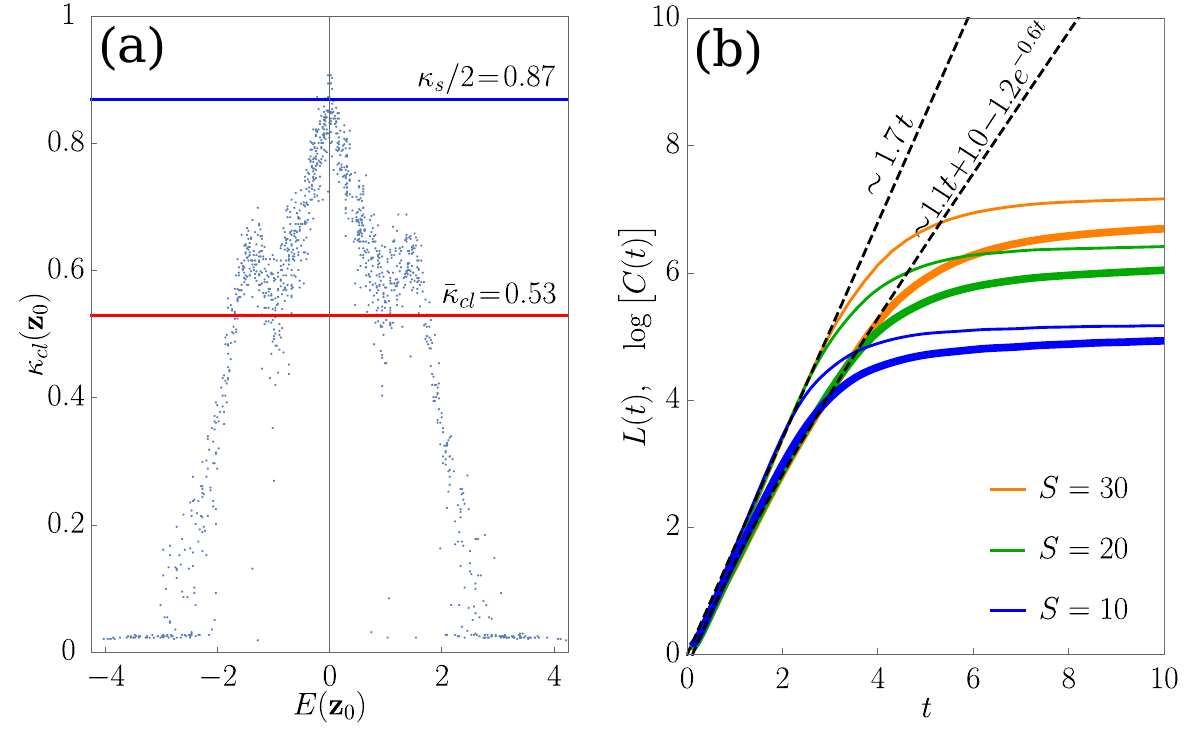}
    \caption{(a) Classical LEs \textit{vs.} energies for 1250 randomly generated initial conditions in the chaotic classical FP model. (b) Infinite-temperature OTOCs (thin lines) and LOTOCs (thick lines) in the quantum FP model.}
    \label{fig:FP}
\end{figure}

\textit{True chaos}.-- To study the behavior of the OTOC and the LOTOC in a truly chaotic system, we consider the Feingold-Peres (FP) model~\cite{Feingold:1983, Feingold:1984, Fan:2017}:
\beq \label{eq:FP}
\hat{H}_\mathrm{FP} = \hat{x}_1 + \hat{x}_2 + 4 \hat{z}_1 \hat{z}_2, \eeq
where $(\hat{x}_i, \hat{y}_i, \hat{z}_i)$ are independent rescaled $SU(2)$ spin operators. In the classical limit $S \to \infty$, this model has positive LEs for the majority of initial conditions, see Fig.~\ref{fig:FP}(a). In other words, the phase-space average of classical LEs $\bar \kappa_{cl} \approx 0.53 > 0$, so this system is considered classically chaotic. Moreover, model~\eqref{eq:FP} has two isolated saddle points $x_1 = x_2 = \pm 1$, at the vicinity of which $\pd z_i(t) / \pd z_j(0) \sim e^{\kappa_s t}$ with $\kappa_s = \sqrt{3}$. We emphasize that $\max\!\left[\kappa_{cl} \right] \approx \kappa_s / 2 > \bar{\kappa}_{cl}$ because ``overly chaotic'' regions take only a small fraction of the phase space.

The numerical calculation, Fig.~\ref{fig:FP}(b), confirms the qualitative difference between truly chaotic systems and integrable systems with isolated saddle points. In a chaotic system, \textit{both} OTOC and LOTOC grow according to a chaotic pattern until the ``chaotic'' Ehrenfest time: $C(t) \sim e^{2 \kappa_q t}$ and $L(t) \approx 2 \bar{\kappa}_q t$ for $1 \lesssim t \lesssim \log(1/\hbar)$. Furthermore, the LOTOC reproduces the average classical LE, $\bar{\kappa}_q \approx \bar{\kappa}_{cl}$, whereas the OTOC grasps only the contribution from the saddle points, $\kappa_{q} \approx \kappa_{s}/2$. This again confirms that the LOTOC correctly describes the semiclassical behavior of a quantized Hamiltonian system.

Another prominent example of a truly chaotic system is the quantized Arnold cat map~\cite{Esposti, Hannay, Garcia-Mata:2018, Moudgalya}. In this model, the LOTOC also correctly reproduces the Ehrenfest time, $t_E \sim \log(1/\hbar)$, and the classical LE, $\bar{\kappa}_q \approx \bar{\kappa}_{cl}$, see~\cite{companion}.

\textit{Many-body chaos}.-- To illustrate the replica trick~\eqref{eq:replica-trick}, we consider the system of $N \gg 1$ nonlinearly coupled oscillators, which is inspired by the spatial reduction of the $SU(2)$ Yang-Mills model~\cite{Matinyan, Chirikov, Savvidy:1984, Savvidy:2022}:
\beq \label{eq:H-ON}
\hat{H} = \left[ \frac{1}{2} \hat{p}_i^2 + \frac{1}{2} m^2 \hat{x}_i^2 + \frac{\lambda}{4 N} \hat{x}_i^2 \hat{x}_j^2 \right] -\frac{\lambda}{4 N} \hat{x}_i^4, \eeq
where we assume the summation over the repeated indices and single out the $O(N)$-symmetric part. The classical counterpart of this model is chaotic for $N \ge 2$, and the average classical LE is estimated as $\bar \kappa_{cl} \approx 0.7 \sqrt[4]{\lambda T}/N$ in the large-$N$ and high-temperature limit~\cite{Kolganov}.

\begin{figure}
    \centering
    \includegraphics[width=\linewidth]{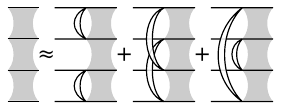}
    \caption{The approximate Dyson-Schwinger equation that sums the leading exponentially growing contributions to $C_2(t)$ in model~\eqref{eq:H-ON}. Horizontal lines denote retarded propagators on different folds of the Keldysh contour, and vertical crescents denote resummed bubble chains that connect symmetric and nonsymmetric vertices on different folds.}
    \label{fig:DS-ON}
\end{figure}

To estimate replica OTOCs, we write down the tree-level correlation functions on the $2n$-fold Keldysh contour and resum the loop corrections. The leading in~$1/N$ corrections preserve the $O(N)$ symmetry; so, in this approximation, model~\eqref{eq:H-ON} is approximately integrable, and replica OTOCs simply oscillate with time. Nevertheless, the next-to-leading order in $1/N$ contains the contributions from the nonsymmetric vertices that modify the Dyson-Schwinger equation on the resummed replica OTOC. Substituting an exponentially growing ansatz $C_n(t) \sim e^{2 n \varkappa_n t}$ into this equation, we reduce it to the equation on $\varkappa_n$:
\beq \label{eq:ON-kappa-eq}
1 \approx (2n-1)!! \left[ -\frac{1536}{N^2} \frac{\lambda^2}{(\tilde{\mu} \tilde{m})^6} \frac{e^{\tilde{\beta} \tilde{m}}}{\big( e^{\tilde{\beta} \tilde{m}} - 1 \big)^2} \frac{\tilde{m}^4}{\big( \tilde{m}^2 + \varkappa_n^2 \big)^2} \right]^n.
\eeq
The leading contribution to this equation is ensured by the ``mixed multi-rung'' ladder diagrams (Fig.~\ref{fig:DS-ON}). For brevity, we introduce short notations for the inverse temperature of the replicated model $\tilde{\beta} = (n+1)/T$, the resummed mass $\tilde{m}$, and the parameter of the resummed vertex $\tilde{\mu}$. The solution to Eq.~\eqref{eq:ON-kappa-eq} gives an approximate expression for the replica LE:
\beq \label{eq:ON-kappa}
\kappa_n = n \varkappa_n \approx n \left[(2n-1)!!\right]^{\frac{1}{2n}} \frac{8 \sqrt{6}}{N} \frac{\lambda \tilde{m}}{(\tilde{\mu} \tilde{m})^3} \frac{e^{\tilde{\beta} \tilde{m}/2}}{e^{\tilde{\beta} \tilde{m}} - 1}. \eeq

Finally, employing the replica trick~\eqref{eq:replica-trick}, we estimate the refined quantum LE in the high-temperature and weak-coupling limit, $m/T \ll \lambda/m^3 \ll 1$:
\beq \label{eq:ON-true-kappa} 
\bar{\kappa}_q \approx 0.7 \sqrt[4]{\lambda T}/N. \eeq

\begin{figure}
    \centering
    \includegraphics[width=\linewidth]{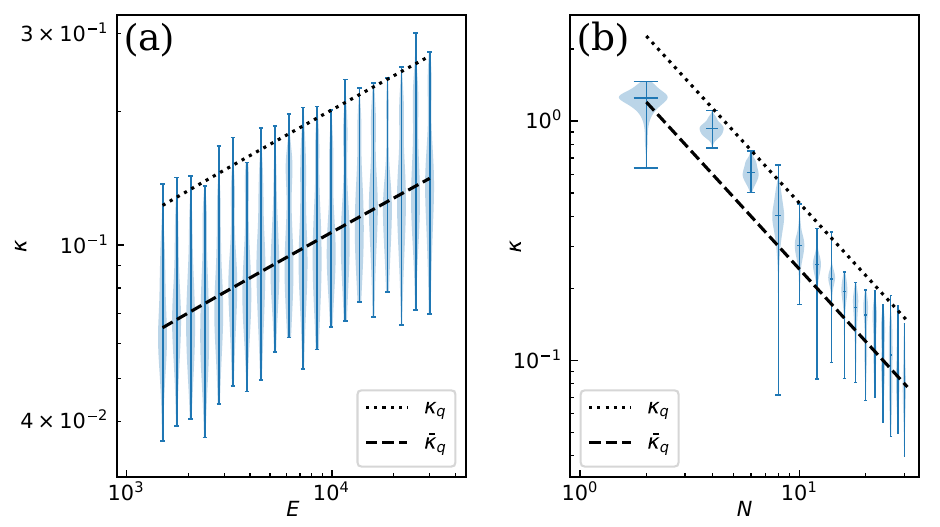}
    \caption{Numerically calculated classical LEs $\kappa_{cl}$ (vertical bars and violins), conventional quantum LEs $\kappa_q$, and refined quantum LEs $\bar{\kappa}_q$ for a fixed number of oscillators $N = 30$ (a) or energy $E = 100 m^4/\lambda$ (b) in model~\eqref{eq:H-ON}.}
    \label{fig:ON}
\end{figure}

We emphasize that the refined quantum LE is approximately two times smaller than the conventional LE, $\kappa_q \approx 1.3 \sqrt[4]{\lambda T}/N$. From the diagrammatic perspective, the refined quantum LE is reduced by correlations between different replicas, which do not factorize and leave footprints in the behavior of the LOTOC (Fig.~\ref{fig:DS-ON}). From the semiclassical perspective, the discrepancy arises because classical LEs have a nontrivial distribution in the phase space (Fig.~\ref{fig:ON}). The LOTOC measures the fair average of LEs, so $\bar{\kappa}_q \approx \bar{\kappa}_{cl}$ as $\hbar \to 0$; on the contrary, the OTOC selects only the points with the largest LEs, so $\kappa_q \approx \max\left[\kappa_{cl}(\mathbf{z}_0)\right] > \bar{\kappa}_{cl}$ as $\hbar \to 0$, where the maximum is taken over all initial conditions $\mathbf{z}_0$. In this respect, model~\eqref{eq:H-ON} is similar to the FP model, where the average and maximum classical LEs also differ, cf. Fig.~\ref{fig:FP}(a).

\textit{Maximal chaos}.-- Maximally chaotic quantum systems, where OTOCs exponentially grow with time and saturate the bound~\cite{MSS}, are especially notable for their putative duality to black holes. The Sachdev-Ye-Kitaev (SYK) model is probably the most prominent example of such a system. This is a quantum mechanical model of $N$ Majorana fermions $\chi_i$ with all-to-all random couplings~\cite{Polchinski, Maldacena-SYK, Kitaev, Sarosi, Rosenhaus, Trunin-SYK}:
\beq \label{eq:SYK}
\hat{H}_\mathrm{SYK} = i^{q/2} \sum_{1 \le k_1 \le \cdots \le k_q \le N} j_{k_1 \cdots k_q} \chi_{k_1} \cdots \chi_{k_q}, \eeq
where numbers $j_{k_1 \cdots k_q}$ are drawn from a Gaussian distribution with zero mean and the following variance:
\beq \label{eq:SYK-J}
\left\langle j_{k_1 \cdots k_q}^2 \right\rangle = \frac{2^{q-1}}{q} \frac{J^2 (q-1)!}{N^{q-1}} \quad \text{(no sum)}. \eeq

\begin{figure}
    \centering
    \includegraphics[width=\linewidth]{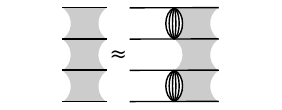}
    \caption{The approximate Dyson-Schwinger equation that sums the leading exponentially growing contributions to $C_2(t)$ in the SYK model with $q=8$. Horizontal lines denote retarded propagators, and vertical lines denote Wightman propagators that connect different folds of the Keldysh contour.}
    \label{fig:DS-SYK}
\end{figure}

Let us calculate the refined quantum LE of the SYK model and show that it saturates the bound~\cite{MSS} similarly to the conventional quantum LE. For simplicity, we consider the limit $N \gg q \gg 1$, where the leading contributions to the exact propagators are calculated explicitly. To estimate these contributions, we first solve the Dyson-Schwinger equation on the Euclidean propagator that lives on the imaginary-time part of the $2n$-fold Keldysh contour. In the limit $N \gg q \gg 1$, this equation resums the melonic diagrams and has the following approximate solution:
\beq \label{eq:SYK-propagator}
G(\tau) \approx \frac{1}{2} \mathrm{sgn}(\tau) \left[ 1 + \frac{2}{q} \log \frac{\cos \frac{\pi \tilde{v}}{2}}{\cos \left( \frac{\pi \tilde{v}}{2} - \frac{\pi \tilde{v} |\tau|}{\tilde{\beta}} \right)} \right], \eeq
where the parameter $\tilde{v}$ is determined from the equation $ \tilde{\beta} J = \pi \tilde{v} / \cos\left(\pi \tilde{v}/2\right)$ and $\tilde{\beta} = (n+1)/T$ is the inverse temperature of the replicated model. Analytically continuing propagator~\eqref{eq:SYK-propagator} to real times, we obtain all propagators on the $2n$-fold Keldysh contour. Then, similarly to model~\eqref{eq:H-ON}, we write down the Dyson-Schwinger equation on the resummed replica OTOC (Fig.~\ref{fig:DS-SYK}), substitute the ansatz $C_n(t) \sim e^{2 n \varkappa_n t}$, and obtain the equation on~$\varkappa_n$:
\beq \label{eq:SYK-kappa-eq}
1 \approx \left[ 2 \pi \tilde{v} / \tilde{\beta} \varkappa_n \right]^n. \eeq
Note that Eqs.~\eqref{eq:ON-kappa-eq} and~\eqref{eq:SYK-kappa-eq} have different combinatorial prefactors due to the peculiarities of the large-$N$ limits in models~\eqref{eq:H-ON} and~\eqref{eq:SYK}. Finally, solving Eq.~\eqref{eq:SYK-kappa-eq} and substituting the solution into Eq.~\eqref{eq:replica-trick}, we obtain the refined quantum LE in the limit in question:
\beq \label{eq:SYK-kappa}
\bar{\kappa}_q = 2 \pi T v, \eeq
where the parameter $v$ is determined from the equation $J/T = \pi v/\cos(\pi v/2)$.

We emphasize that in the SYK model, refined and conventional quantum LEs coincide because correlations between different folds of the Keldysh contour are suppressed by the powers of $1/N$, see Fig.~\ref{fig:DS-SYK}. Indeed, the replica OTOCs simply factorize, $C_n(t) \sim \left[ C_1(t) \right]^n$, so the replica trick~\eqref{eq:replica-trick} yields $\bar{\kappa}_q = \lim \varkappa_n = \kappa_q$. In particular, this implies that the refined quantum LE~\eqref{eq:SYK-kappa} saturates the bound $\bar{\kappa}_q \le 2 \pi T$ in the low-temperature limit $T \ll J$.

\textit{A bound on refined chaos}.-- One of the most important achievements of OTOCs is the bound on quantum~LE~\cite{MSS}, $\kappa_q \le 2 \pi T$, which resolves the cloning paradox~\cite{Hayden, Sekino, Lashkari} and helps to find holographic duals of black holes. We argue that the refined quantum LE also satisfies this bound.

First, the semiclassical picture discussed in the previous sections implies that the refined quantum LE coincides with the phase-space \textit{average} of classical LEs, whereas the conventional quantum LE approaches the phase-space \textit{supremum} of classical LEs as $\hbar \to 0$. Since the average cannot be greater than the maximum element of the set, the refined quantum LE is less than or equal to the conventional one. Hence, it automatically satisfies the bound $\bar{\kappa}_q \le \kappa_q \le 2 \pi T$. 

Second, the replica LEs are proved to satisfy the bound $\kappa_n \le 2 \pi T n$ for any positive integer $n$~\cite{Pappalardi, Tsuji}. Analytically continuing this inequality to real $n$, keeping in mind that $\kappa_n \ge 0$ by definition, and employing a variant of the replica trick~\eqref{eq:replica-trick}, $\bar{\kappa}_q \to \left( \kappa_n - \kappa_0 \right) / n$ as $n \to 0$, we straightforwardly obtain the bound $\bar{\kappa}_q \le 2 \pi T$. 

Finally, calculations in the SYK model, which is dual to the nearly-AdS$_2$ gravity with matter~\cite{Maldacena-JT, Jensen, Engelsoy, Gross-1, Gross-2}, show that the refined and conventional quantum LEs can coincide and saturate the bound~\cite{MSS} together. Hence, the saturation of the inequality $\bar{\kappa}_q \le 2 \pi T$ is still a useful indicator of the gauge/gravity duality.

\textit{Discussion}.-- We have shown that the LOTOC correctly describes the semiclassical behavior of quantized Hamiltonian systems. Unlike the conventional OTOC, it correctly reproduces the Ehrenfest time and the average classical LE in both chaotic and integrable systems, including systems with isolated saddle points. Of course, our case studies are by no means exhaustive. In particular, it is very interesting to examine the behavior of the LOTOC after the Ehrenfest time. However, our examples make it sufficiently clear that the LOTOC provides a proper definition of quantum chaos and quantum butterfly effect --- the exponential sensitivity to \textit{typical} small perturbations ensured by $\bar{\kappa}_q > 0$, --- and correctly reproduces the definition of classical chaos in the limit $\hbar \to 0$.

At the same time, the LOTOC retains all advantages of the OTOC. In particular, it can be calculated analytically employing the replica trick and the extended Schwinger-Keldysh diagram technique, which are especially useful for large-$N$ quantum systems dual to black holes. Moreover, the refined quantum LE satisfies the fundamental bound on chaos, $\bar{\kappa}_q \le \kappa_q \le 2 \pi T$. So, in a sense, our approach reconciles the definitions of quantum chaos and scrambling separated by the observations of~\cite{Xu}. Besides, the LOTOC and the replica OTOCs can be experimentally measured using the same protocols as conventional OTOCs~\cite{Garcia-Alvarez, Jafferis:Nature, Garttner, Li, Green}. Thus, the LOTOC fixes the flaws of conventional OTOCs, where these flaws are important, and has a comparably wide range of applications, from thermalization of quantum systems to teleportation through a traversable wormhole (e.g., see~\cite{Jafferis:Nature, Yoshida, Gao:2019}).

\textit{Acknowledgments}.--
We thank Nikita Kolganov and Artem Alexandrov for the collaboration at the initial stage of this project. We are also grateful to Anatoly Dymarsky, Elizaveta Trunina, Andrei Semenov, Vladimir Losyakov, Petr Arseev, Damir Sadekov, and Alexey Rubtsov for valuable discussions. The classical LEs at Fig.~\ref{fig:FP}(a) were calculated using package~\cite{Sandri}. This work was supported by the Foundation for the Advancement of Theoretical Physics and Mathematics ``BASIS''.

\bibliography{refs}

\end{document}